\documentclass[5p,twocolumn,authoryear]{elsarticle}
\usepackage{url}

\journal{Journal of \LaTeX\ Templates}
\begin{document}
\begin{frontmatter}
\title{IAU Meteor Data Center --- the shower database: a~status report}


\author[mymainaddress]{Tadeusz Jan Jopek}

\author[mysecondaryaddress]{Zuzana Ka\v{n}uchov\'{a}\corref{mycorrespondingauthor}}

\cortext[mycorrespondingauthor]{Corresponding author}
\ead{zkanuch@ta3.sk}

\address[mymainaddress]{Institute Astronomical Observatory, Faculty of Physics, A.M. University, Pozna\'{n}, Poland}
\address[mysecondaryaddress]{Astronomical Institute of the Slovak Academy of Sciences, 05960 Tatransk\'{a} Lomnica, Slovakia}

\begin{abstract}
Currently, the meteor shower part of Meteor Data Center database includes: 112 established showers,  563 in the working list, among them 36 have the {\it pro tempore} status. The list of shower complexes contains 25 groups, 3 have established status and 1 has the {\it pro tempore} status. 

In the past three years, new meteor showers submitted to the MDC database were detected amongst the meteors observed by CAMS  stations (Cameras for Allsky Meteor Surveillance), those included in the EDMOND (European viDeo MeteOr Network Database), those collected by the Japanese SonotaCo Network, recorded in the IMO (International Meteor Organization) database, observed by the Croatian Meteor Network and on the Southern Hemisphere by the SAAMER radar. 

At the~XXIXth General Assembly of the~IAU in Honolulu, Hawaii in 2015, the names of 18 showers were officially accepted and moved to the list of established ones. Also, one shower already officially named (3/SIA the Southern iota Aquariids) was moved back to the working list of meteor showers.
At the XXIXth GA IAU the basic shower nomenclature rule was modified, the new formulation predicates ``The general rule is that a meteor shower (and a meteoroid stream) should be named after the constellation that contains the nearest star to the radiant point, using the possessive Latin form".

Over the last three years the MDC database was supplemented with the earlier published original data on meteor showers, which permitted verification of the correctness of the MDC data and extension of bibliographic information.
Slowly but surely new database software options are implemented, and software bugs are corrected. 
\end{abstract}

\end{frontmatter}


\section{Introduction}
The Meteor Dat{\bf{a}} Center shower database (MDC) came into being in 2007 \citep{Jenniskens2008, JopekJenniskens2011, JopekKanuchova2014} as a complement to the already existing orbital part of the IAU Meteor Data Center \citep{Lindblad1987, Lindblad1991, LindbladSteel1994, Lindblad2003, Svoren2008, Porubcan2011, Neslusan2014}. 

The main objective of the MDC database is to archive information on meteor showers: their geocentric and heliocentric parameters. Before inclusion into the MDC, a new meteor shower obtains a unique name and code, according to a set of officially formalized rules. One of the results of this procedure is a considerably less confusion in the meteor shower names published in literature.

The MDC showers database is posted on the Web on two servers: at the Astronomical Observatory, A.\,M. University in Pozna\'{n}, Poland (\url{http://localhost/~jopek/MDC2007/}) and at the Astronomical Institute of the Slovak Academy of Sciences, 05960 Tatransk\'{a} Lomnica, Slovakia (\url{https://www.ta3.sk/IAUC22DB/MDC2007/}).

Beyond the efficient collection,  checking and dissemination of the meteor showers data, the main task of the MDC (in the  conjunction with the Working Group on Meteor Shower Nomenclature of IAU Commission F1) is  to formulate a descriptive list of the established meteor showers that can receive official names during the upcoming IAU General Assembly. 

The first Working Group (named at that time the Task Group) was appointed at the XXVIth GA IAU in 2006 held in Prague. The activity of each WG lasts three years in the period between two IAU General Assemblies.
At present, starting from  the XXIXth GA IAU in Honolulu, the current members of the WG in the 2015-2018 triennium are: Diego Janches (chair), Peter Brown, Peter Jenniskens, Tadeusz J. Jopek (editor of the MDC shower database), Zuzana Ka\v{n}uchov\'{a}, Gulchekhra I. Kokhirova, Masahiro Koseki, Regina Rudawska, Josep M. Trigo-Rodriguez, and Jun-Ichi Watanabe.

In the following sections, we give some information about the MDC structure, about the meteor shower data included in the database, and we discuss some problems encountered in the archiving process. Finally,  we address to some future plans related with MDC. 
\section{Content of the lists of meteor showers}
At the time of the Meteoroids 2013 meeting in Pozna\'{n} the IAU MDC database stored the data of 579 showers \citep{JopekKanuchova2014}. The list of established meteor showers contained 95 records, the working list included 460 meteor showers, among them 95 had {\it pro tempore} status. The list of shower groups contained 24 complexes, three of them had the established status.  
 At present 700 showers are archived and their parameters are grouped into five listings.
\begin{enumerate}
\item {\bf List of all showers}. It contains actually 700 showers. These are all showers (except the removed ones) registered in the Database. 
\item {\bf List of established showers} contains 112 ones. A shower can be moved from the Working list to the list of established showers if: 1) it passes the verification procedure and 2) is given  an official name which has to be approved by the IAU F1 Commission business meeting during the next GA IAU. 
 
Exceptionally, the established meteor shower may be moved back to the working list (see section \ref{newestablished}).

In the time period between the last and the next GA IAU one can nominate a shower from the working list to become an established shower (its IAU numerical code appears with a green background in the list). The list of nominated candidates is discussed and completed shortly before the GA IAU and it is proposed for the approval by the adequate Commission during the GA IAU. 
\item {\bf The Working list} contains 563 showers that were already published in the scientific literature or in the IMO WGN Journal and 36 new submissions with the ``pro tempore" flag (the IAU numerical code is with the orange background). The ``{\em pro tempore}" flag is removed after publication (or acceptation for publication). The working list showers can stay on this list for many years, until they fulfill verification criteria and obtain official names accepted at the GA IAU. 

If a  new submission is not be published within ~2 years, it is removed permanently from the database.
A shower can be removed from the database also on the basis of the published analysis.

\item {\bf List of shower groups} contains actually 25 shower complexes. Three of them have the status ``established", one group obtained  ``{\em pro tempore}" flag.
\item {\bf List of removed showers}. In the past some showers were included in the working list but for some reasons they were removed from it. Because  these showers can be found in literature, we think that we should  keep complete information about the contents of the database. The List of Removed Showers is a kind of historical archive. 
Actually 10 showers are in this list. They were removed from the Working list for different reasons. In most cases a given shower had been proved to be already observed and registered in MDC. 
\end{enumerate}
The content of all lists mentioned above may be displayed by the Web browser, four of them can be downloaded as ASCII files.
 
As mentioned above, the list of all showers contains the data on 700 showers. However, 228 of them are represented by two or more sets of radiants and orbital parameters. Additional sets of parameters were collected by us from the meteor literature, but some of them originate from recent observations and were submitted by the MDC users. The shower MDC database contains inclusively 1184 data records. Multiple sets of shower parameters play important functions, apart from their cognitive importance, each independent set of parameters strongly confirms the existence of a particular meteor shower.
\section{Newly established showers}
\label{newestablished}
 Meteor showers were officially named for the first time in the history of astronomy at the XXVIIth GA IAU held in Rio de Janeiro in August 2009. At this conference 64 meteor showers were given official names \citep{JopekJenniskens2011}. At the following XXVIIIth GA IAU in Beijing 2012 further 31 official names of the meteor showers were approved \citep{JopekKanuchova2014}. 
Other 18 showers were named officially in 2015 at the XXIXth GA IAU in Honolulu. Their geocentric and heliocentric parameters are given in Table 1.  

Thirteen of the showers that have been named recently have radiants located on the northern hemisphere, most likely because of the distribution of meteor observing sites on the Earth.  About half of the newly established showers moves on prograde orbits. Only five showers have near ecliptic orbits. Particles of eleven showers {collide} with the Earth at the geocentric velocities greater than $50$\,km/s. The shower 510/JRC June rho Cygnids has an exceptional mean orbit that is perpendicular to the plane of the ecliptic.

At the GA IAU in Honolulu, at the request of Peter Jenniskens, the Working Group on Meteor Shower Nomenclature proposed to move the established meteor shower 3/SIA (the Southern iota Aquariids) back to the working list \citep[for more details see][]{Jenniskens2015}. Identification of a reliable meteor shower is a quite complex task. The new results are confronted with those obtained in the past, which, due to the complex structure of the meteoroid streams and their variable dynamical evolution, hinders finding a final solution. Therefore, switching of the meteor shower status should be considered as something inevitable. Fortunately, such changes happen only exceptionally.  
\begin{table*}
\begin{center}
\label{tabela1}
\caption{Geocentric radiants and heliocentric orbital data of 18 meteor showers (streams) officially named pending the~ \mbox{XXIXth} GA IAU held in Honolulu in 2015. 
The~solar ecliptic longitude $\lambda_{S}$ at the time of shower maximum activity, the~geocentric radiant right ascension and declination $\alpha_g$, $\delta_g$ and the~values of the~angular orbital elements $\omega, \Omega, i$ are given for the~epoch J2000.0. Numerical values of the meteoroid parameters are the average values determined from N individual sets of radiants and orbits.
\vspace{0.1cm} }
\footnotesize
\begin{tabular*}{\textwidth}{cclrrrrrrrrrrr}
\hline
 \multicolumn{2}{c}{IAU}     &   \multicolumn{1}{c}{Shower  name}      & \multicolumn{1}{c}{$\lambda_{S}$} & \multicolumn{1}{c}{$\alpha_g$} & \multicolumn{1}{c}{$\delta_g$}  & \multicolumn{1}{c}{$V_g$} & \multicolumn{1}{c}{$a$} & \multicolumn{1}{c}{$q$} & \multicolumn{1}{c}{$e$} & \multicolumn{1}{c}{$\omega$}  & \multicolumn{1}{c}{$\Omega$} & \multicolumn{1}{c}{$i$} &\multicolumn{1}{c}{$N$}   \\
 \multicolumn{1}{r}{No} & \multicolumn{1}{l}{Code}   & \multicolumn{1}{l}{} & \multicolumn{1}{c}{$\deg$} & \multicolumn{1}{c}{$\deg$} & \multicolumn{1}{c}{$\deg$} & \multicolumn{1}{c}{km/s} & \multicolumn{1}{c}{au}& \multicolumn{1}{c}{au} & \multicolumn{1}{c}{ }& \multicolumn{1}{c}{$\deg$}& \multicolumn{1}{c}{$\deg$}  & \multicolumn{1}{c}{$\deg$} & \multicolumn{1}{c}{ }  \\ 
\hline  
21 &  AVB &alpha Virginids                    & 32.0& 203.5 &   2.9&  18.8 & 2.55  & 0.744 & 0.716  & 274.9 & 30.0  &   7.0 & 12 \\  
69 &  SSG &Southern mu Sagittariids           & 86.0& 273.2 & -29.5&  25.1 & 2.02  & 0.457 & 0.769  & 104.5 & 266.4 &   6.0 & 70 \\
96 &  NCC &Northern delta Cancrids            &296.0& 127.6 &  21.5&  27.2 & 2.23  & 0.410 & 0.814  & 286.6 & 290.0 &   2.7 & 74 \\
97 &  SCC &Southern delta Cancrids            &296.0& 125.0 &  14.4&  27.0 & 2.26  & 0.430 & 0.811  & 105.0 & 109.3 &   4.7 & 69 \\ 
343&  HVI &h Virginids                        & 38.0& 204.8 & -11.5&  17.2 & 2.28  & 0.742 & 0.659  &  72.7 & 218.2 &   0.9 & 11 \\
362&  JMC &June mu Cassiopeiids               & 74.0& 17.5  &  53.9&  43.6 &57.24  & 0.577 & 0.990  & 97.68 & 74.0  &  68.3 & 584\\ 
428&  DSV &December sigma Virginids           &262.0&200.8  &  5.8 &  66.2 & 8.18  & 0.565 & 0.971  &  97.9 & 261.8 & 151.5 & 22 \\
431&  JIP &June iota Pegasids                 & 94.0& 332.1 & 29.1 &  58.5 & 7.44  & 0.903 & 0.928  & 219.9 & 94.1  & 112.8 & 11 \\
506&  FEV &February epsilon Virginids         &314.0& 200.4 & 11.0 &  62.9 & 8.28  & 0.491 & 0.954  & 272.5 & 312.6 & 138.0 & 55 \\
510&  JRC &June rho Cygnids                   & 84  & 321.8 & 43.9 &  50.2 & 21    & 1.007 & 0.931  & 190   &  84.2 & 90    & 16 \\
512&  RPU &rho Puppids                        &231.0&  130.4& -26.3&  57.8 & 9.40  & 0.987 & 0.915  & 349.4 &  50.8 & 107.0 & 22 \\
524&  LUM &lambda Ursae Majorids              &215.0&  158.2&  49.4&  60.3 & 13    & 0.917 & 0.931  & 147   & 215.0 & 115   & 29 \\
526&  SLD &Southern lambda Draconids          &221.6&  163  &  68.1&  48.7 & 4.0   & 0.986 & 0.744  & 189   & 221.6 & 88.0  & 26 \\
529&  EHY &eta Hydrids                        &256.9&  132.9&  2.3 &  62.5 & 15    & 0.383 & 0.974  & 103.8 &  76.9 & 142.8 & 120\\
530&  ECV &eta Corvids                        &302  &  192.0& -18.1&  68.1 & 5.29  & 0.820 & 0.847  &  50.1 & 122.2 &  50.1 & 16 \\
533&  JXA &July xi Arietids                   &119  &   40.1&  10.6&  69.4 &  -    & 0.883 & 0.965  & 318   & 299   &  171.6& 61 \\
549&  FAN &49 Andromedids                     &112.0&   20.5&  46.6&  60.2 & 7.71  & 0.898 & 0.922  & 139.8 & 118.0 & 117.9 & 76 \\
569&  OHY &omicron Hydrids                    &309  &  176.3& -34.1&  59.1 &  -    & 0.684 & 0.931  &  68.6 & 128.9 & 114.3 & 29 \\
\hline
\end{tabular*}  
\normalsize 
\end{center}
\label{kuku}
\end{table*}
\section{Modification of the meteor shower nomenclature rules}
The traditional meteor shower nomenclature was formalized by adopting a set of rules \citep{Jenniskens2007, Jenniskens2008, JopekJenniskens2011}. The basic rules predicated {\em ``The general rule is that a meteor shower should be named after the constellation of stars that contains the radiant \ldots"} and also {\em ``to distinguish among showers from the same constellation the shower may be named after the nearest (brightest) star \ldots"}. However, sometimes such definitions  are not sufficient from the practical point of view. 
One can see in Fig.~1 that the radiant point of January nu Hydrids is placed very close to the border between the Hydra and Sextans constellations. With ``naked eye" it is not possible to decide which constellation this radiant lies in. In another case the radiant clearly lies in a given constellation, but the nearest stars belongs to another one, see Fig.~1.
\begin{figure}[t!]
\begin{center} 
\vbox{
\includegraphics[width=0.72\linewidth]{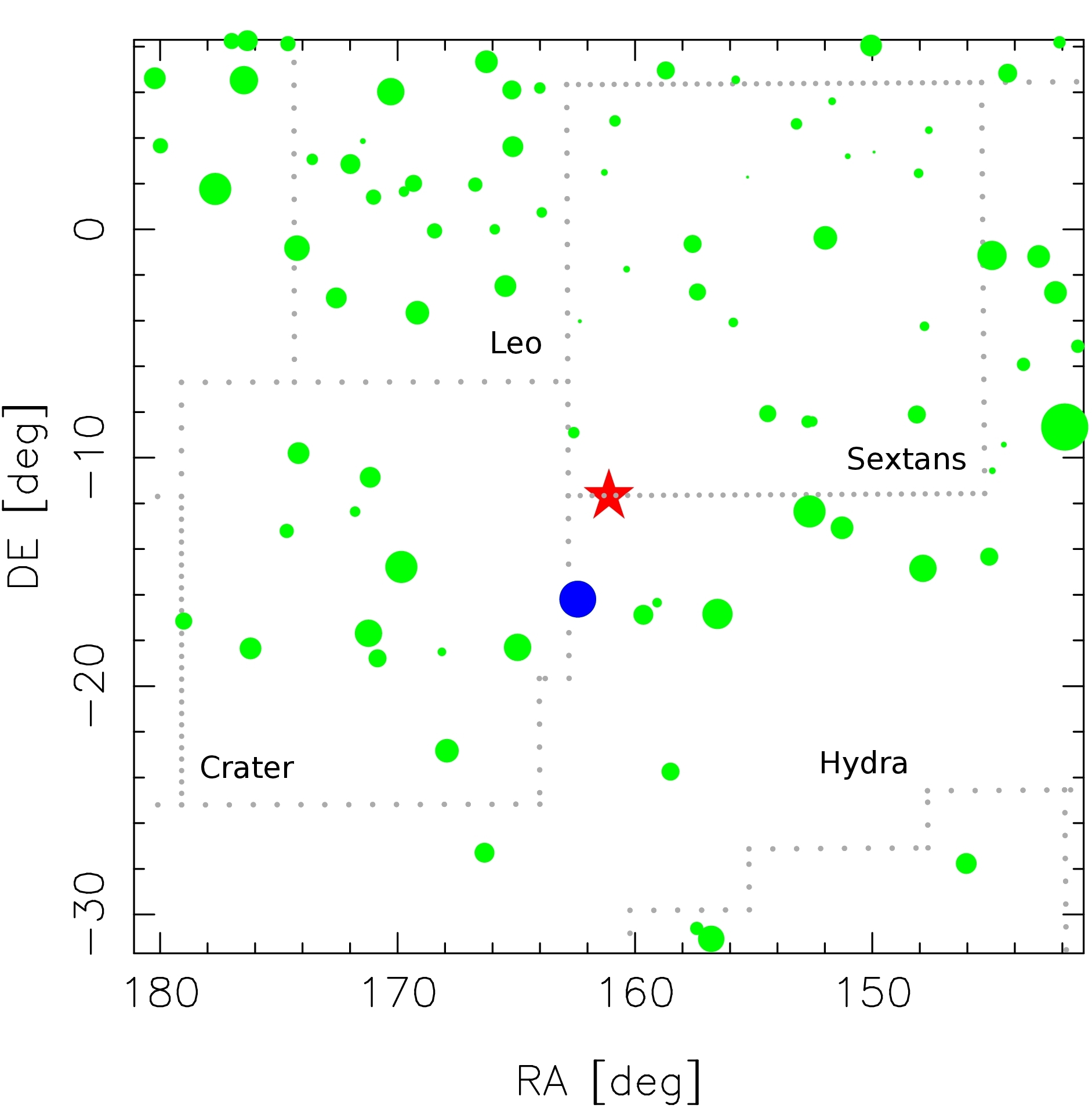}\\
\vspace{0.4cm}
\includegraphics[width=0.72\linewidth]{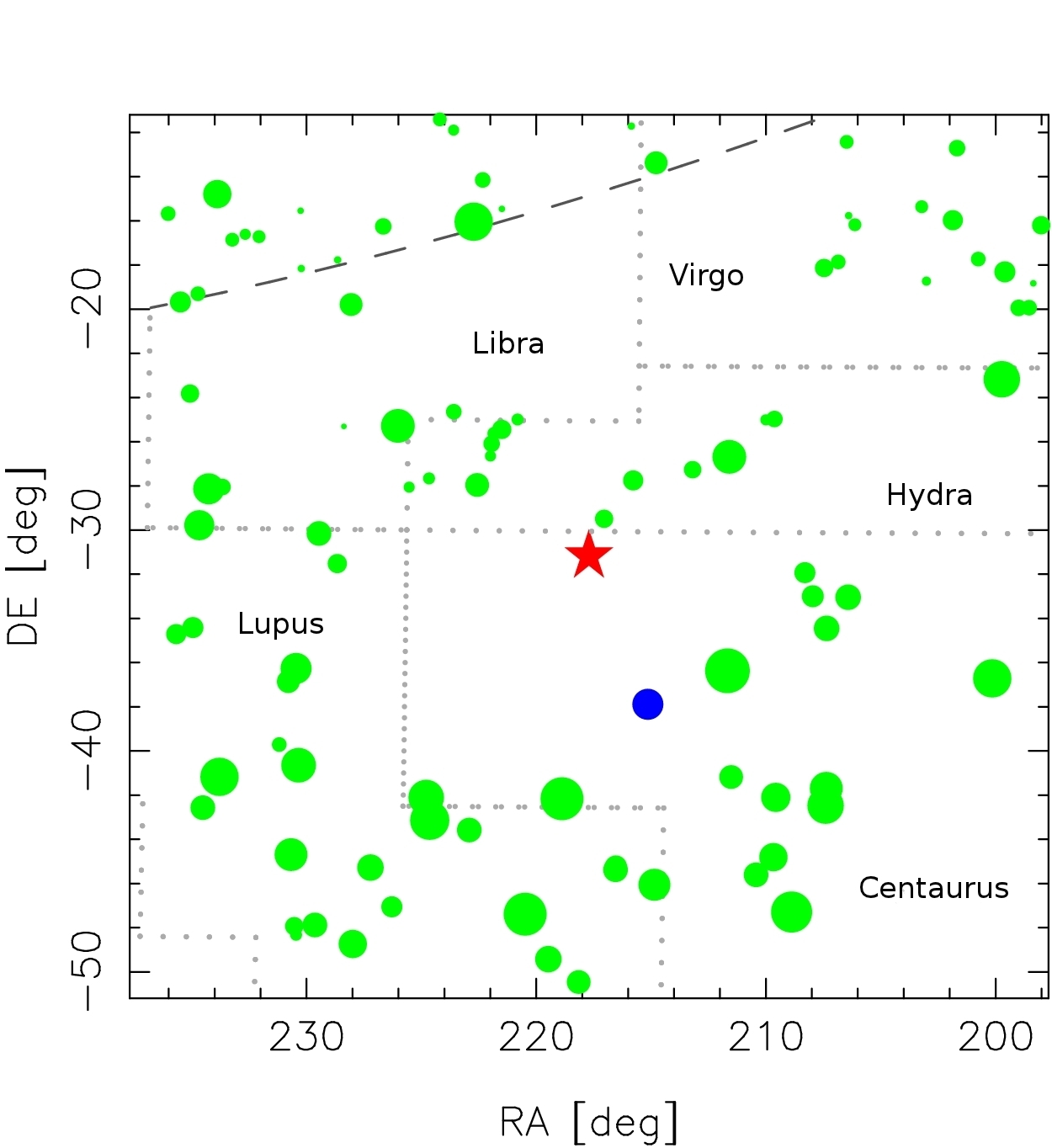}
\vspace{0.4cm}
}
\caption{Top panel: the difficult case of the~radiant placement (marked by a star symbol) of January nu Hydrids (544/JNH). {The radiant} lies almost at the~border of two constellations. Bottom panel: {The radiant of }the May lambda Virginids (148/MLV) lies in the Centaurus constellation, but the closest star lies in  the Hydra constellation. This shower was named before any shower nomenclature rules were formalized. 
}
\end{center}
\label{rikitiki}
\end{figure}
For these reasons, the current wording in the nomenclature rule reading: ``The
general rule is that a meteor shower (and a meteoroid stream) should be named after
the current constellation that contains the radiant, specifically using the possessive
Latin form" should be replaced with a new rule reading: ``The general rule is that a meteor shower (and a meteoroid stream) should be named after the constellation that contains the nearest
star to the radiant point, using the possessive Latin form".
This nomenclature change was adopted during the~IAU Commission 22 business meeting in Helsinki at the ACM 2014 conference. It was clarified in the discussion that the nearest star with a Bayer designation, a Greek or Roman letter, or (in exceptional cases) Flamsteed number is meant. 
This nomenclature change was approved by voting at the XXIXth IAU GA in Honolulu.
\section{Errors, mistakes, shortcomings and positives}
The MDC shower database is certainly not perfect --- it contains some erroneous data, mistakes and various shortcomings. The reasons for this state are many varying from our distraction to insufficient experience. However nobody is perfect and the erroneous data can be also found in the literature from which we derived the meteor shower data. 

The meteor parameters given in the MDC are not homogeneous.  They originate from different epochs and they  are determined from visual, photographic, video and radar observations. Thus the uncertainties of the parameters are different. There is another problem (also noticed by \cite{Andreic2014}).  Some authors supply to the MDC only geocentric parameters without orbital information. Furthermore, sometimes apparent radiant parameters are submitted instead of the geocentric ones. It is not always possible to recognize the difference reading the scientific source literature. Similarly, reading literature, quite often one cannot find any information about the method used for averaging of the meteor shower parameters. Different methods result in different mean values of the radiant and the orbital parameters. Thus, often inconsistencies are found between the values of the semi-major axis, eccentricity, perihelion distance and the argument of perihelion of the meteoroid stream orbit given in MDC. 

Hence, despite our efforts, MDC contains the erroneous data and mistakes. Therefore with a view to improve the database we really appreciate every critical remark related to the information archived in MDC.
  
The MDC database is maintained on a voluntary basis, which limits our ability to handle complex submissions and to develop user interface utilities.

We appreciate very much all positive signals about the MDC shower database, e.g.  ``You might be interested to know that colleagues and I are using the information you sent me to estimate the location and intensity of meteoroid streams striking the lunar surface. This will be used to help with the interpretation of observations from the upcoming LADEE mission:  \url{http://www.nasa.gov/mission_pages/LADEE/main/index.html#.Ue1uT43VCSo}.", \citep{Timoty2013}.
We found the other two in \cite{Andreic2014}:  ``The most valuable option of the IAU MDC web page is the possibility to download the current shower list \ldots" and  ``Nice touch is that the shower data frequently include the link to the reference from which the data originate". We are very grateful for all such encouraging opinions.
\begin{figure}[t!]
\begin{center} 
\vbox{
\includegraphics[width=0.72\linewidth]{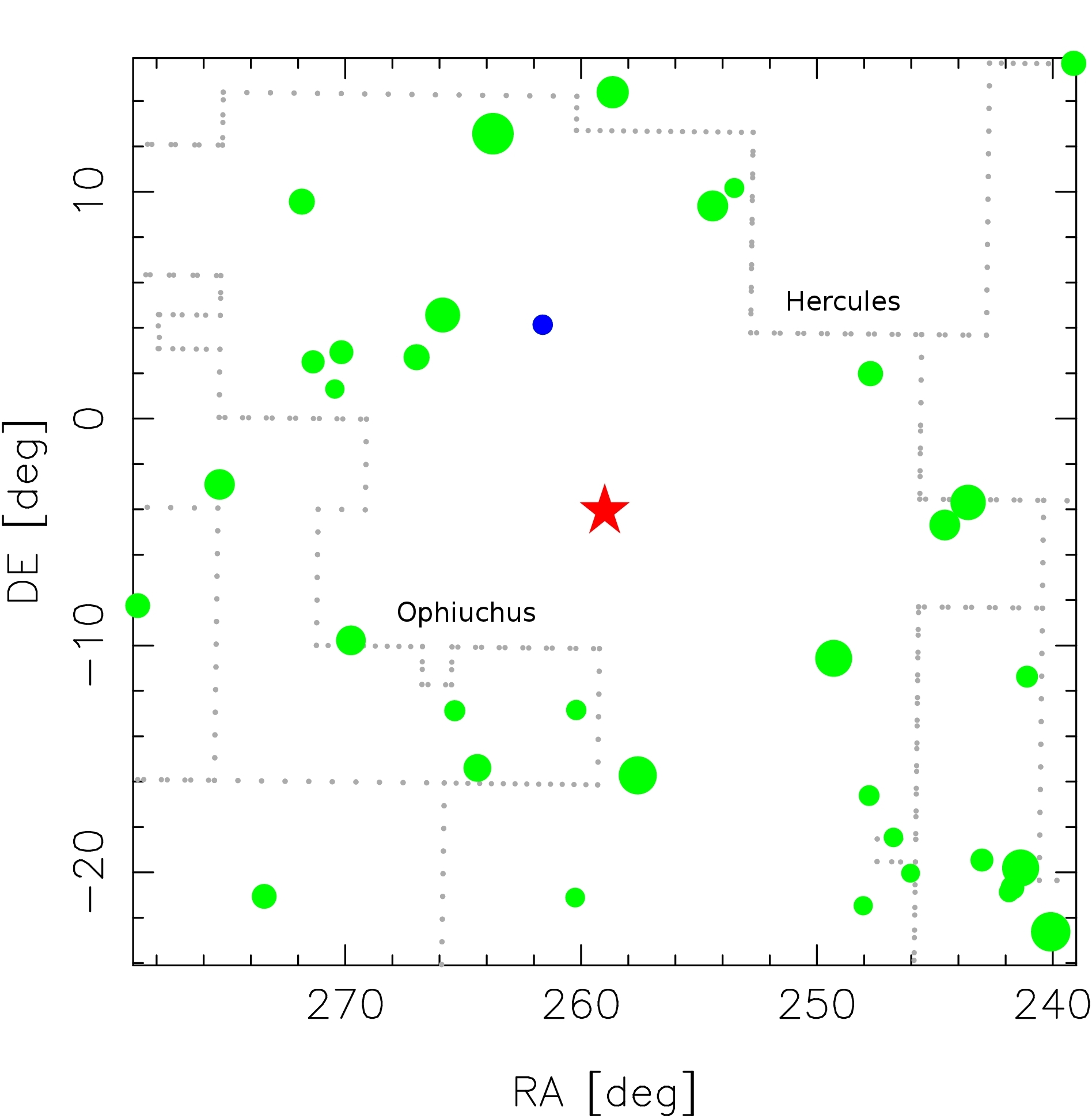}\\
\vspace{0.4cm}
\includegraphics[width=0.72\linewidth]{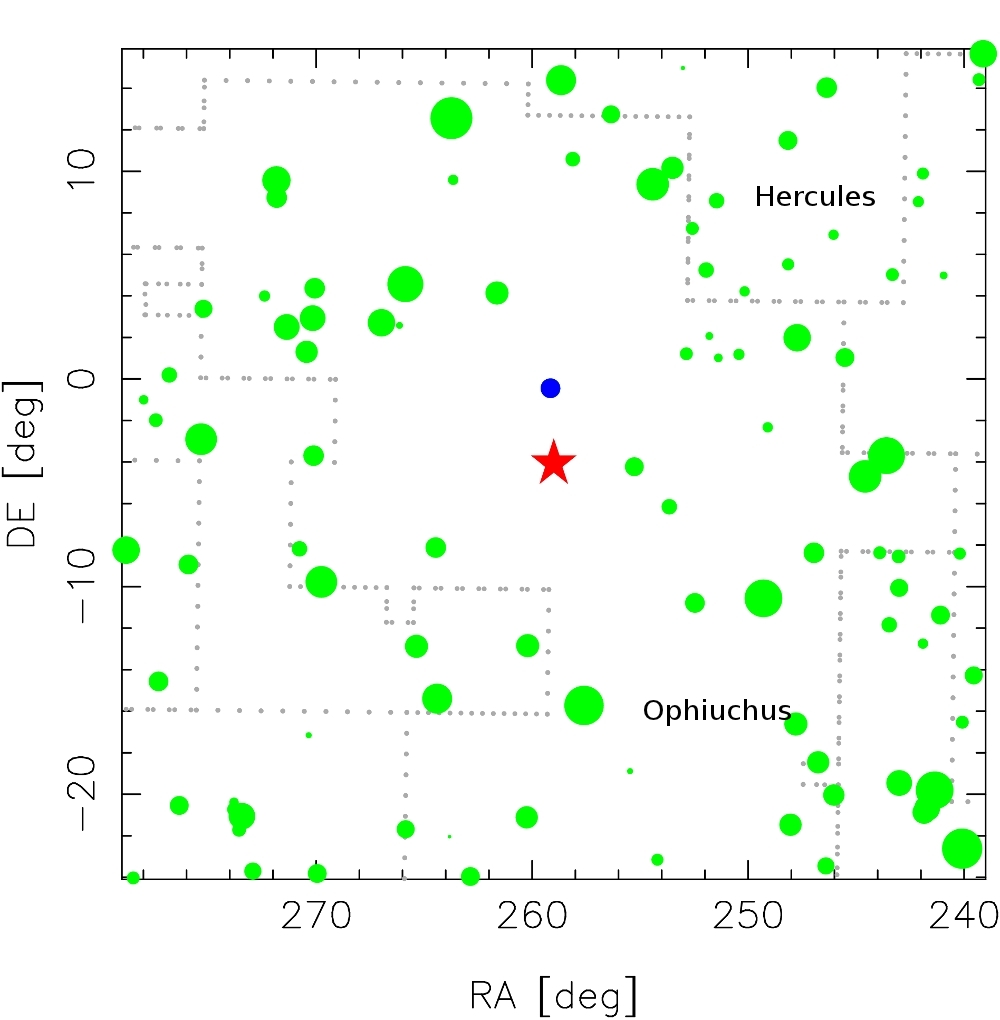}
\vspace{0.4cm}
}
\caption{The radiant of  the meteor shower 536/FSO, 47 Ophiuchids, (marked by  a star symbol) plotted on two star maps: at the top panel, the stars were taken from the subset of the PPM star catalogue; and at the bottom panel, the stars originate from the BSC catalogue. The nearest star to the radiant is in both cases marked by a blue circle. As one can see, adjudication of the name of this shower  depends on the selection of the catalogue. 
}
\end{center}
\label{rura}
\end{figure}
\section{Further MDC shower database activity}
%
At first, we have to continue our main task which is the assignment of IAU  codes to the  new meteor showers submitted to MDC.  

As far as possible, the MDC shower database will be improved. As first we will complete archiving of the original source meteor data found in literature. Also we will complete collecting the hyper-links to the ADS meteor shower literature references.

Following the remark given by \cite{Andreic2014}, we will add the periods of activity of the meteor showers to the database.
We also plane to implement a software which will ease  the employment of the MDC database.

Despite the recent modification of the meteor shower nomenclature rules, we suggest to the Working Group on Meteor Shower Nomenclature to accept the Yale Bright Star Catalogue (BSC), 5th Revised Ed. \citep{Hoffleit1991} as the standard for the naming of the meteor showers.

The reason for such standardization is clearly seen in Fig.~2. The top panel was plotted using the subset of $\sim 900$ stars taken from the Positions and Proper Motions Star Catalogue \citep{Roeser1991}. The bottom map is plotted using stars from the Yale Bright Star Catalogue containing $3141$ stars. 
As shown adjudication of a shower name depends on the choice of the nearest star to the meteor radiant and thus it depends on the selection of stars in the star catalogue. 

Finally, we plan to move more showers to the List of Removed Showers if such recommendation is published in literature. 
\section*{Acknowledgments}
We would like to acknowledge the~users of the~IAU MDC who informed us about several erroneous data: \v{Z}eljko Andrei\'{c}, Regina Rudawska and Damir \v{S}egon.  \v{Z}eljko Andrei\'{c}, Damir \v{S}egon and Denis Vida are acknowledged for their excellent  ``A statistical walk through the IAU MDC database'' \citep{Andreic2014}. Z.K. was supported by VEGA - The Slovak Agency for Science, Grant No. 2/0032/14.

The authors acknowledge both anonymous referees for their useful comments.

This research has made use of NASA's Astrophysics Data System Bibliographic Services.
\section*{References}

\end{document}